\documentclass{JHEP3} 

\usepackage{epsfig}

\newcommand\fverb{\setbox\pippobox=\hbox\bgroup\verb}
\newcommand\fverbdo{\egroup\medskip\noindent%
                        \fbox{\unhbox\pippobox}\ }
\newcommand\fverbit{\egroup\item[\fbox{\unhbox\pippobox}]}
\newbox\pippobox
\title{Towards the solution of noncommutative $YM_2$: Morita equivalence
and large N-limit}

\author{ L. Griguolo \\
Dipartimento di  Fisica, Universit\`a  di Parma,
INFN-Gruppo collegato di Parma\\
Parco Area delle Scienza 7/A, 43100 Parma, Italy\\
E-mail: \email{griguolo@fis.unipr.it}}
\author{D. Seminara and P. Valtancoli\\
Dipartimento di Fisica, Polo Scientifico Universit\'a di Firenze,
INFN Sezione di Firenze\\
Via  G. Sansone 1, 50019 Sesto Fiorentino, Italy\\
Email: \email{seminara@fi.infn.it, valtancoli@fi.infn.it}
}

\received{\today}               
\accepted{\today}               
\preprint{UPRF-2001-22}      

\date{data}
\abstract{In this paper we shall investigate the possibility of  solving
$U(1)$ theories on the non-commutative (NC) plane for arbitrary values
of $\theta$ by exploiting Morita equivalence. This duality maps the NC
$U(1)$ on the   two-torus  with a rational parameter $\theta$ to the
standard $U(N)$  theory in the presence of a 't Hooft flux, whose
solution is completely known. Thus, assuming a smooth dependence on
$\theta$, we are able to construct a series rational approximants
of the original theory, which is finally reached by taking the large
$N-$limit at fixed 't Hooft flux. As we shall see, this procedure
hides some subletities since the approach of $N$ to infinity is linked
to the shrinking of the commutative two-torus to zero-size. The volume
of NC torus instead diverges and it provides a natural cut-off for some
intermediate steps of our computation. In this limit,
we shall compute both the partition function and the correlator of
two Wilson lines. A remarkable fact is that the configurations, providing
a finite action in this limit, are in correspondence with
the non-commutative solitons (fluxons) found independently by Polychronakos
and by Gross and Nekrasov, through a direct computation on the plane.}

\keywords{Noncommutative Gauge Theories, Wilson loop, Large N-limit}

\begin{document}

\section{Introduction}
\indent
\ \ \ \ \ \
Non-commutative field theories have gained a central role in the recent
developments of string theory. The initial interest was motivated by the
results presented in \cite{cds}, where non-commutative geometry was found
to be the natural tool  to classify new toroidal compactifications of
$M-$theory in the presence of a constant background three-form field.
Later on, in \cite{sw}, non-commutative gauge theories were shown to appear
in $IIA/B$ superstring theory  in a particular decoupling limit of
$D-$branes with a $NS-NS$ two-form background turned on.
This possibility to embed
consistently a non-commutative field theory into a string theory has
stimulated a large amount of studies, trying to understand
classical and quantum non-commutative dynamics both at perturbative \cite{mrs}
and at nonperturbative \cite{gn} level.

Non-commutative field  theories (and in particular gauge theories)
present a large variety of new phenomena not completely
understood even in the basic cases: at perturbative level the $UV/IR$
mixing\footnote{We give here only the references where the phenomenon was
discovered in four-dimensional scalar and gauge theories.}
\cite{mrs,suski} complicates the renormalization program
(see however \cite{gp} for a recent discussion of renormalization in
non-commutative QFT) and it seems to produce tachyonic instabilities
\cite{espe}. At the same
time an entirely new family of classical solutions has been discovered
\cite{ns}, and their role in the quantum dynamics is completely unknown (see
however \cite{valya} where the effects of instantons in
$N=2$ non-commutative SUSY gauge theory were considered). Moreover,
gauge theories on the non-commutative torus
exhibit a fascinating property, not shared by their commutative
ancestors, that goes under the name of Morita equivalence. Roughly speaking,
Morita equivalence establishes a relation between gauge theories defined on
different non-commutative tori: gauge theories characterized by
diverse ranks of the gauge group, flux numbers and  non-commutative
parameters are seen to be equivalent \cite{schwarz}.
This beautiful and absolutely general mathematical property has an elegant
and simple interpretation when a string theory embedding of the field
theoretical model is available and the non-commutative torus
originates from  a compactification procedure. In fact, in this case,
Morita equivalence can be viewed as a consequence of the more
familiar $T$-duality \cite{dual}. Nevertheless, we must stress again that it
holds without any reference to string theory and it is a general
(nonperturbative) property, depending just on the geometrical data
of the theory itself.

This deep geometrical origin suggests that Morita equivalence must
play a central role in understanding some basic facts on non-commutative
gauge theories both on $T^D$ and on $I\!\!R^D$ (the non-commutative
$I\!\!R^D$ can be recovered as a suitable large volume limit of a
non-commutative torus $T^D$). To this purpose, a particular promising
feature is that, under a Morita transformation, a $U(1)$ gauge theory on
a non-commutative torus of ${\it rational}$ parameter
\footnote{In the following we shall use a bidimensional language since we
 shall deal with only this case. In higher dimensions the parameter $\theta$,
for  example, is substituted by an antisymmetric matrix $\theta_{\mu\nu}$}
$\theta$ is shown to be equivalent to a certain $U(N)$ Yang-Mills theory,
in the presence of a `t Hooft flux \cite{baal}, defined on a ${\it commutative}
$ torus. This is not case when $\theta$ is irrational (see also \cite{boris}
for a discussion on the phase structure of ``irrational'' theories).
Although the question of the smooth dependence on $\theta$ of the theory
is still under investigation \cite{gura,luis}, this unexpected link opens
the concrete possibility to study a non-commutative  gauge theory for a
generic value of the non-commutative parameter starting from a series
of commutative approximants \cite{Lizzi}.
What we have in mind is, of course, to define a limiting procedure for the
commutative theory in order to reach a general non-commutative parameter on the
non-commutative Euclidean space. We did not attempt to do this in four
dimensions: we choose the simpler two-dimensional case, where a complete
understanding of the commutative theory is available \cite{migdal}.
At the same
time the general solution of non-commutative $U(1)$ theory in $D=2$ does not
exist, while previous studies on classical solutions \cite{poly,gn2} and Wilson
loops \cite{bnt} show that highly non-trivial aspects are involved. In this
paper we propose a limiting procedure to construct $U(1)$ theory on the
non-commutative plane, for general $\theta$, starting from a two-torus of
rational parameter: it is a particular large $N$-limit at a fixed value of the
't Hooft flux. This limit was also recently considered in \cite{luis,gura2},
where, however, the four dimensional case was under study.

In this limit some unexpected, but promising effects emerge. Since the
't Hooft flux is kept fixed, the large volume limit (encoded in the large
$N$-limit) corresponds in the commutative approximants to the shrinking of the
torus to zero-size. This make our analysis more involved, because the
behavior of $YM_2$ on a small torus is quite delicate \cite{transi}.
At the same time, the volume of the non-commutative torus diverges and
it becomes the natural cut-off for the intermediate steps of the
computations. Moreover, in this procedure,
some contributions are
naturally singled out. We shall call them {\it finite action configurations}
\footnote{The origin of this name will become manifest in sec. 3}. It is
easy to see that they are in correspondence with the classical
solutions carrying a non trivial flux (fluxons) discussed by Polychronakos
\cite{poly} and  by Gross and Nekrasov
\cite{gn2}. Under the semiclassical assumption that these are the only
configurations dominating our limit, we obtain an exact partition function
which has the nice property to be extensive. We must recall that, in the
commutative case the semiclassical approximation is exact \cite{witten}
and this may suggest a more general validity of our results.

The setting for computing correlators of Wilson lines is also presented.
The complete computations is carried out in the case of two lines and
its semiclassical   interpretation in terms of fluxons is given.
This technique allows us to tackle also the computations of others
observables such as closed Wilson loops and their correlators. This
question as well as the one on the exactness of the semiclassical
assumption is under investigation at the present \cite{gsv}.

\section{Non-commutative $U(1)$ theory on the torus and Morita equivalence}

Gauge theories on the non-commutative plane can be constructed by replacing,
in the usual Yang-Mills action, the ordinary commutative product of functions
with the Moyal $\star$-product. Its definition is given by
\begin{equation}
f(x)\star g(x)=\exp\left(i\frac{\theta^{\mu\nu}}{2}
\frac{\partial}{\partial x^\mu}
\frac{\partial}{\partial y^\nu} \right)\left. f(x)g(y) \right|_{y\to x}.
\end{equation}
Notice that, endowed with this product, the commutative coordinates $x^\mu$
are promoted to  satisfy the familiar Heisenberg algebra
\begin{equation}
\label{ncx}
x^\mu\star x^\nu-x^\nu\star x^\mu=i~\theta^{\mu\nu}.
\end{equation}
This signals that noncommutativity will, in general, violate Lorentz
invariance.
However, in two dimensions, which is the case
under investigation,  the  relation (\ref{ncx}) does not break Lorentz
(or Euclidean) invariance, since~ $\theta^{\mu\nu}$ can be always expressed
in term  of the invariant Levi-Civita tensor,
\begin{equation}
\theta^{\mu\nu}=\theta~ \epsilon^{\mu\nu}.
\end{equation}

\noindent
The action of the non-commutative $U(N)$ Yang-Mills theory can be thus written
in this general form
\begin{equation}
S=\frac{1}{4g^2} \int d x^2
{\rm Tr}\Bigl[(F_{\mu\nu}+\Phi_{\mu\nu}I\!\!I)\star(F^{\mu\nu}+
\Phi^{\mu\nu}I\!\!I)\Bigr],
\end{equation}
where
\begin{equation}
F_{\mu\nu}(x)=\partial_{\mu} A_\nu-\partial_\nu A_\mu-i(A_\mu\star A_\nu
-A_\nu\star A_\mu)
\end{equation}
and $\Phi_{\mu\nu}=\Phi\epsilon_{\mu\nu}$ is a $U(1)$ background term
whose meaning will become clear shortly. We notice that a new dimensional
parameter, $\theta$, has been introduced into the theory through the
star-product, having exactly the same dimensions of the space-time
($[length]^2$), a point that will be appreciate later.

\noindent
When we compactify both coordinates to go from the plane to the torus,
this theories exhibit an $SO(2,2,Z\!\!\!Z)$ Morita equivalence which is
inherited from string theory T-duality \cite{dual}. The very same property
has also been demonstrated explicitly without recourse to either
string theory or supersymmetry \cite{schwarz}.
The equivalence connects different
non-commutative gauge theories living  on different non-commutative tori:
in fact the duality group has an $SL(2,Z\!\!\!Z)$ subgroup which acts as
follows on the geometrical and gauge data
\begin{equation}
\left (\begin{array}{c} m' \\ N'\end{array}\right)=
\left( \begin{array}{ccc}a & & b\\ c & & d\end{array}\right)
\left (\begin{array}{c} m \\ N\end{array}\right), \ \ \ \ \
\Theta'=\frac{c+d\Theta}{a+b\Theta}
\label{fon1}
\end{equation}
\begin{equation}
(R')^2=R^2(a+b\Theta)^2,\ \ \ \ (g')^2=g^2|a+b \Theta|
\ \ \ \  \tilde{\Phi}'=(a+b\Theta)^2\tilde{\Phi}-b(a+b\Theta)
\label{fon2}
\end{equation}
where $\Theta\equiv\theta/(2\pi R^2)$, $\tilde{\Phi}=2\pi R^2\Phi$
and $(2 \pi R)$ is the circumference of the torus, which for simplicity
we take to be square. The first entry, $m$, in the column-vector denotes
the magnetic flux, while $N$ is the rank of the gauge group.
The parameter of the transformation are integer numbers, constrained
by the equation $ad-bc=1$. Let us notice that changing the overall
sign of the matrix realizing the $SL(2,Z\!\!\!Z)$ transformation leads to
the same $\Theta'$. The sign has to be chosen so that the new gauge
group has positive rank.
An important observation, that will be
crucial for our development, is that the $\theta$ parameter, defining the
Moyal product, scales as the area of the torus, while Morita equivalence only
involves the adimensional quantities $\Theta$. Another important point is
that the background connection $\Phi$ transforms inhomogeneously under Morita,
therefore it is, in general, non vanishing along the $SL(2,Z\!\!\!Z)$ orbit.
Let
us now consider a $U(1)$ gauge theory with ${\it rational}$ $\Theta=-c/N$
(with $gcd(c,N)=1$), magnetic flux $m_{nc}$ and, for simplicity, vanishing
background connection: a $SL(2,Z\!\!\!Z)$ transformation of the form
\begin{equation}
\label{segno}
M=
\left( \begin{array}{ccc}a & & b\\ c & &
N\end{array}\right),
\end{equation}
brings the theory to $\Theta=0$, giving therefore an ordinary but nonabelian
theory. In eq. (\ref{segno}) we choose the global sign of $c$ and $N$ and
consequently that of $a$ and $b$ so that the new rank of the gauge group
is positive\footnote{The sign ambiguity in $M$ can be in fact simply encoded in
the following observation: the couples $(c,N)$ and $(-c,-N)$ give the same
$\Theta$. The change sign of the couple $(a,b)$ is then constrained by the
Diophantine relation $a~ N-c~ b=1$. To be precise, this relation determines $(a,b)$ only
up to a solution of the homogeneous equation  $a~ N-c~ b=0$, but this  does
not alter our conclusion.}. Working out the effect of $M$ via eq.(\ref{fon1})
we see that the new gauge group is
\begin{equation}
\label{Nc}
N_c=c~m_{nc}+N.
\end{equation}
The magnetic flux is also changed, and we may express the parameter $b$ as
a linear combination of the noncommutative magnetic flux and the commutative
one
\begin{equation}
b=m_c           -a~m_{nc}.
\end{equation}
Here $m_c$ is the magnetic flux (the ``commutative'' flux). The parameter $a$
and $m_c$ are then constrained by the ${\rm det} M=1$ condition that is
\begin{equation}
a~ N-c (m_c-a~ m_{nc})=a~ N_c-c~ m_c=1.
\label{ANA}
\end{equation}
This Diophantine equation determines $a$ modulo $c$ and $m_c$ modulo $N_c$.
An important point in our analysis will be to show that these ambiguities
does not affect our limiting procedure.
A background connection has also to be introduced into the theory,
according to eq. (\ref{fon2}), leading to
\begin{equation}
\tilde{\Phi}_c=-\frac{b}{N}=-\frac{m_c~-a~m_{nc}}{N}.
\label{beck}
\end{equation}
Our original non-commutative theory is therefore equivalent to an ordinary
theory on a torus with area shrunk by a factor $N^2$
\begin{equation}
R^2_c=\frac{R_{nc}^2}{N^2}.
\end{equation}
The coupling constant has also become weaker of a factor $N$:
$g^2_c=g^2_{nc}/N$.
Thus one could take the point of view that a non-commutative theory
exists at least for rational $\Theta$, and then try to define
the theory at irrational values by approaching it with an infinite sequence
of rational numbers. This possibility has been advocated a certain number of
times in the literature \cite{gura,luis,gura2,bigatti}, but,
at least at our knowledge,
non
concrete computation have been proposed in two dimensions.
We take here a slightly different point of view in that we want to construct the non-commutative theory on the non-commutative plane with arbitrary finite $\theta$ parameter: we intend therefore to perform the limit $R_{nc}\to\infty$ (decompactifying the original
non-commutative torus where the non-commutative $U(1)$ theory was defined),
maintaining finite the dimensional parameter
$\theta$\footnote{After having completed the computations the very same
proposal appears in \cite{luis,gura2}}. We recall in fact that
on the plane $\theta$ has dimension $[length]^2$, that is
not related to any obvious geometrical quantity. Because we have
\begin{equation}
\label{segno1}
\theta=2\pi R^2_{nc}\Theta=-2\pi R^2_{nc}\frac{c}{N},
\end{equation}
an easy way to realize our task is to work with a radius
\begin{equation}
R^2_{nc}=-N\frac{\theta}{2\pi c},
\label{dec}
\end{equation}
and taking the limit $N\to\infty$. The value of $\theta$ can now be taken
completely general, therefore we can freely choose $c$ since this will
correspond to a rescaling of the final $\theta$. The Morita equivalent
theory with $\theta=0$ is therefore defined by the following data
\begin{equation}
R^2_c=-\frac{\theta }{2\pi}\frac{1}{c~ N}= \frac{|\theta|}{2\pi |c ~N|} ,
\label{defi1}
\end{equation}
\begin{equation}
g_c^2=\frac{g_{nc}^2}{N}.
\label{defi2}
\end{equation}
The last equality in (\ref{defi1}) is a consequence of the fact that
$\theta$ and $\Theta$ are taken to have the same sign because of
(\ref{segno1}). Next we have to satisfy $SL(2,Z\!\!\!Z)$ constraint:
\begin{equation}
m_c=\frac{1}{c}(a~ N_c-1)\,\,\,\,\,
\label{defi3}
\end{equation}
It may seem strange, at this level, that the commutative Chern class is not fully determined
(we have the arbitrary integer $a$ in its definition), but we will see how its
ambiguity does not affect the final result.
Notice that as $N$ goes to infinity, for any fixed non-commutative flux, the
Chern class of the equivalent commutative gauge theory has to scale
accordingly. Eqs. (\ref{defi1}, \ref{defi2}, \ref{defi3}) define the limit we
would like to perform: it is clear that this is not the usual 't Hooft limit
on a commutative torus. Eq. (\ref{defi2}) alone would define the usual
large
$N$-limit: in addition eq. (\ref{defi1}) means that we have also to perform
a small area limit and eq. (\ref{defi3}) forces us to push the first Chern
number to infinity. Moreover we have to consider the effect of the $U(1)$
background connection, that will exactly compensate the arbitrariness of $a$.
What we are going to do in the next section, is to use
the exact solution of the Yang-Mills theory on the torus \cite{migdal} in
order to construct the Morita equivalent theory for finite $N$, to implement
the presence of the flux and of the background connection, and eventually to
perform our large $N$-limit.

\section{The partition function of the Morita equivalent theory and its
large $N$-limit}

\noindent
The 't Hooft limit of two-dimensional Yang-Mills theory is a well studied
problem: in particular Gross and Taylor has shown \cite{w4} that when the
space-time is a compact Riemann surface a string theory emerges in the
large $N$-limit. The dependence of the theory on the size of the area has
also been investigated: on the sphere it does exist a critical area separating
a strong coupling phase from a weak coupling phase \cite{DK}. The phase
transition does not occur for genus greater than one: on the torus it would
happen just at zero area. The small area behavior on the torus has been
studied in \cite{ruud}, in connection with the string description: in our
case we have to study a similar limit, with the area scaling as $1/N$.
No one, instead, has considered the possibility to perform also a limit on
the first Chern class, that is a peculiar element in the Morita equivalent
description.

\noindent
In the following we will consider the case of a $U(1)$ non-commutative theory
with vanishing first Chern class ($m_{nc}=0)$ and vanishing background connection
($\Phi_{nc}=0$). We leave open for future investigations the
possibility to sum over the different non-commutative Chern classes.

\noindent
We start by considering the $U(N)$ theory defined by the following action
\begin{equation}
S=\frac{1}{4g_c^2} \int d x^2
{\rm Tr}\left[\left(F_{\mu\nu}-\frac{m}{2\pi R^2_c N}
\epsilon_{\mu\nu}I\!\!I\right)\left(F^{\mu\nu}-
\frac{m}{2\pi R^2_c N}\epsilon^{\mu\nu}I\!\!I\right)\right],
\label{actio}
\end{equation}
where the explicit form of the background connection,
$\Phi_c=-\frac{m}{2\pi R^2_c N}I\!\!I$ has been taken into
account. The Chern class of the $U(N)$ field is $m$
\begin{equation}
m=\frac{1}{4\pi}\int d x^2
{\rm Tr}[F_{\mu\nu}\epsilon^{\mu\nu}].
\end{equation}
It is therefore possible to work out the $\Phi$-dependence
in eq. (\ref{actio})
\begin{equation}
S=\frac{1}{4g_c^2} \int d x^2
{\rm Tr}[F_{\mu\nu}F^{\mu\nu}]-
\frac{2\pi^2  m^2}{g^2_cA_c N},
\label{actio1}
\end{equation}
where $A_c=4\pi^2R^2_c$ is the area of the commutative torus. We have now to
compute the partition function of a $U(N)$ theory with Chern class equal to
$m$ and the additional (constant) term present in eq. (\ref{actio1}). Let us
start with the general Migdal's formula for $U(N)$ (on genus one)
\begin{equation}
\label{partition}
{\cal Z}=\sum_{R} \exp\left[-{{g_c^2 A_c}\over 2}C_2(R)\right],
\end{equation}
where $C_2(R)$ is the value of the second Casimir operator in the
representation $R$. The sum runs over the irreducible representation of the
gauge group: in the $U(N)$ case the
representations $R$ can be labeled by a set of integers
$n_i=(n_1,...,n_N)$, related to the Young tableaux, obeying the ordering
$+\infty>n_1>n_2>..>n_N>-\infty$. In terms of $n_i$ we have for the
second Casimir
\begin{equation}
\label{casimiri}
C_2(R)=C_2(n_1,..,n_N)
=\frac{N}{12}(N^2-1)+\sum_{i=1}^{N}(n_{i}-\frac{N-1}{2})^2.
\end{equation}
The dependence on the product $g_c^2A_c$ is peculiar of two dimensional
Yang-Mills theories, that are invariant under area-preserving
diffeomorphisms. Using the permutation symmetry we get
\begin{equation}
\label{partitione}
{\cal Z}=\frac{1}{N!}\sum_{n_1\neq n_2\neq..\neq n_N}
\exp\left[-{{g_c^2 A_c}\over 2}\sum_{i=1}^{N}(n_{i}-\frac{N-1}{2})^2\right];
\end{equation}
we have disregarded the overall constant term, present in the Casimir,
linked to a cosmological constant contribution that plays no dynamical role in
this context. To fix the Chern class we factorize the $U(1)$ part: we define
\begin{equation}
n_1=\lambda,\,\,\bar{n}_i=n_{i}-n_1\,\,\,\,i=2,..,N,
\end{equation}
obtaining
\begin{eqnarray}
\label{partitione2}
{\cal Z}=&\displaystyle{\frac{1}{N!}\sum_{\lambda=-\infty}^{+\infty}
\sum_{\bar{n}_i\neq
\bar{n}_j\neq 0}
\exp\left[-{{g_c^2 A_cN}\over 2}\left(\lambda-\frac{N-1}{2}+\frac{1}{N}
\sum_{i=2}^{N}\bar{n}_i\right)^2\right]}\nonumber\\
&\displaystyle{\exp\left[ -{{g_c^2 A_c}\over 2}
\left(\sum_{i=2}^{N}\bar{n}_{i}^2-\frac{1}{N}
(\sum_{i=2}^{N}\bar{n}_i)^2\right)\right]}.
\end{eqnarray}
Next we introduce the identity
\begin{equation}
1=\sum_{l=0}^{N-1}\delta_{N}\left(l+\frac{N(N-1)}{2}-
\sum_{i=2}^N\bar{n}_i\right),
\label{delta}
\end{equation}
where $\delta_N$ is the $N$-periodic delta function, $\delta_N(x)=\frac{1}{N}
\sum_{k=0}^{N-1}\exp[-2\pi i\frac{k}{N}x]$. By using eq. (\ref{delta}) we rewrite our
partition function as
\begin{eqnarray}
\label{partitione3}
{\cal Z}=
&&\!\!\!\displaystyle{\frac{1}{NN!}\sum_{l,k=0}^{N-1}\sum_{\bar{n}_i\neq
\bar{n}_j\neq 0} \!\!\!\!\!\!\exp\left[-{{g_c^2 A_c}\over 2}
\left(\sum_{i=2}^{N}\bar{n}_{i}^2-\frac{1}{N}
(\sum_{i=2}^{N}\bar{n}_i)^2\!\!\!-2\pi i {k}\left(\frac{l}{N}+\frac{(N-1)}{2}-
\sum_{i=2}^N\frac{\bar{n}_i}{N}\right)\!\!\right)\!\!\right]}\nonumber\\
&&\displaystyle\left(\sum_{\lambda=-\infty}^{+\infty}
\exp\left[-{{g_c^2 A_cN}\over 2}\left(\lambda+\frac{l}{N}\right)^2\right]
\right).
\end{eqnarray}
In order to single out the contribution of the bundles of Chern class $m$, we
Poisson resum over $\lambda$
\begin{equation}
\displaystyle{\sum_{\lambda=-\infty}^{+\infty}
\exp\left[-{{g_c^2 A_cN}\over 2}\left(\lambda+\frac{l}{N}\right)^2\right]=
\sqrt{\frac{2\pi}{g_c^2A_cN}}\sum_{\lambda=-\infty}^{+\infty}
\exp\left[-\frac{2\pi^2}{g_c^2 A_cN}\lambda^2-2\pi i\lambda\frac{l}{N}\right]}.
\end{equation}
The sum over $l$ can be done giving a $\delta_N(\lambda-k)$, leading to the
factorization
\begin{equation}
\displaystyle{Z=\sum_{k=0}^{N-1}\left(\sqrt{\frac{2\pi}{g_c^2A_cN}}
\sum_{\lambda=k\,mod\,N}
\exp\left[-\frac{2\pi^2}{g_c^2 A_cN}\lambda^2\right]{\cal Z}_k\right)}.
\end{equation}
We have therefore obtained the factorization of the $U(N)$ partition function
according to $U(N)=U(1)\otimes SU(N)/Z_N$: we have that
\begin{eqnarray}
{\cal Z}_k=&&\!\!\!
\displaystyle{\frac{1}{N!}\sum_{\bar{n}_i\neq
\bar{n}_j\neq 0}\exp\left[-\frac{g_c^2 A_c}{2}
\left(\sum_{i=2}^{N}\bar{n}_{i}^2-\frac{1}{N}
(\sum_{i=2}^{N}\bar{n}_i)^2\right)
-\frac{2\pi ik}{N}\left(\frac{N(N-1)}{2}-
\sum_{i=2}^N\bar{n}_i\right)\right]}\nonumber\\
=&&\displaystyle{\sum_{R} \exp\left[-\frac{g_c^2 A_c}{2}C_2(R)\right]
\frac{{\cal X}_R(e^{2\pi i\frac{k}{N}})}{d_R}}.
\end{eqnarray}
${\cal Z}_k$ is easily seen to coincide with the $SU(N)$ partition function
in the $k$-'t Hooft sector \cite{baal}, the sum over the $SU(N)$
irreducible representations $R$ being weighted with the character
${\cal X}_R(e^{2\pi i\frac{k}{N}})$ of the $k$-th $N$-root of the identity.
The partition function of the $U(N)$ theory with first Chern class $m$ is
therefore
\begin{equation}
{\cal  Z}^{(m)}_{U(N)}=\sqrt{\frac{2\pi}{g_c^2A_cN}}
\exp\left[-\frac{2\pi^2}{g_c^2 A_cN}m^2\right]
{\cal Z}_{m}.
\end{equation}
Coming back to eq.(\ref{actio1}) we see that the effect of the
$U(1)$ background connection is simply to cancel the $U(1)$ contribution,
leaving us, finally, with
\begin{eqnarray}
{\cal Z}=&&
\displaystyle{\frac{1}{N!}\sqrt{\frac{2\pi}{g_c^2A_cN}}
\sum_{\bar{n}_i\neq \bar{n}_j}
\int_{0}^{2\pi}\frac{d\alpha}{\sqrt{\pi}}
\exp\left[-(\alpha-\frac{2\pi}{N}\sum_{i=1}^{N}\bar{n}_i)^2-
2\pi i \frac{m}{N}
\left(\frac{(N-1)N}{2}-\sum_{i=1}^{N}\bar{n}_i\right)\right]}\nonumber\\
&&\exp\left[-\frac{g_c^2 A_c}{2}\left(\sum_{i=1}^{N}\bar{n}_{i}^2-
\frac{1}{N}
(\sum_{i=1}^{N}\bar{n}_i)^2\right)\right];
\label{partitione4}
\end{eqnarray}
we will find  useful to have written the partition function in
eq.(\ref{partitione4}) as a sum over $N$ integers, exploiting the procedure
presented in \cite{bgv}. It is intriguing to notice that eq.
(\ref{partitione4}) also
holds when $m_{nc}\ne 0$ when $N$ is identified with $N_c$ given by eq.
(\ref{Nc}) and $m$ with $m_c$ given by eq. (\ref{defi3}).
Due to the effect of the background connection we realize that the
partition function does not depend on the choice of the solution
in eq.(\ref{ANA}): eq.(\ref{partitione4}) depends
on $m_c$ only modulo $N_c$, which is exactly the ambiguity allowed by
(\ref{ANA}). Actually, for $m_{nc}\ne 0$ the cancellation of the $U(1)$
factor is not complete, but  there is a surving contribution that is, however,
only function of the noncommutative geometrical data and thus it is not
affected by the aforementioned  ambiguity. Its explicit expression
is $\displaystyle{\exp\left[-\frac{2\pi^2 N}{g_{nc}^2 A_{nc} N_c}
m_{nc}^2\right]}$.
Coming back to $m_{nc}=0$, we shall use the above fact
to simplify our subsequent analysis. We start choosing $|c|=1$ since this
is only a finite rescaling of $\theta$, then we can set $m=1$
($\theta$ positive) or $m=-1$ ($\theta$ negative) since what we throw
away is zero modulo $N$. We have now to perform
our large $N$-limit in
eq.(\ref{partitione4}): it is convenient to work in the dual representation
obtained by Poisson resumming the series. According to Witten's suggestion
\cite{witten}, in this representation the partition function appears to be
localized around the classical solutions (``instantons''), and the small area
behavior is better understood within this framework.

\noindent
To perform the Poisson resummation we introduce two auxiliary variables
\begin{eqnarray}
{\cal Z}=&&\!\!\!\!
\displaystyle{\frac{1}{N!}\sqrt{\frac{2\pi}{g_c^2A_cN}}
\sum_{\bar{n}_i\neq \bar{n}_j}\!\!
\int_{0}^{2\pi}\!\!\frac{d\alpha}{\sqrt{\pi}}\int_{-\infty}^{+\infty}
\!\!\!\!\!\!dy \int_{-\infty}^{+\infty}\!\!\frac{d\beta}{2\pi}
\exp\left[i\beta(y-\sum_{i=1}^{N}\bar{n}_{i})-
(\alpha-\frac{2\pi}{N} y
)^2\right]}\nonumber\\ &&\displaystyle{\exp\left[-2\pi i
\frac{m}{N} \left(\frac{(N-1)N}{2}-y\right) -\frac{g_c^2 A_c}{2}
\left(\sum_{i=1}^{N}\bar{n}^2_{i}-\frac{y^2}{N}\right)\right]}.
\label{partitione5}
\end{eqnarray}
Next we observe that we can extend the sum over all $\bar{n}_i$ using the
following trick \cite{toro}: our series is of the type
\begin{equation}
\sum_{\bar{n}_i\neq \bar{n}_j} f(\bar{n}_1,..,\bar{n}_N),
\end{equation}
where $f(\bar{n}_1,..,\bar{n}_N)$ is completely symmetric in $\bar{n}_i$.
We can write
\begin{equation}
\sum_{\bar{n}_i\neq \bar{n}_j}\!\! f(\bar{n}_1,..,\bar{n}_N)=
\sum_{\bar{n}_i=-\infty}^{+\infty}\!\!\!\!
\sum_{P}(-1)^{P}\!\!\int_{0}^{2\pi}
\prod_{i=1}^{N}\frac{d\theta_i}{2\pi}\exp\left[-\sum_{j=1}^{N}\theta_{j}(\bar{n}_j-
\bar{n}_{P(j)})\right]f(\bar{n}_1,..,\bar{n}_N),
\label{rapre}
\end{equation}
where no restriction appears on the $\bar{n}_i$'s. $\sum_{P}$ means the
sum over all elements of the symmetric group $S_N$, $P(i)$ denotes the
index $i$ transformed by $P$, while $(-1)^{P}$ is the parity of the
permutation.
One recovers the original form eq.~(\ref{partitione}) by simply integrating
over the angles $\theta_i$ and using the formula
\begin{equation}
\sum_{P}(-1)^{P}\prod_{i=1}^{N}\delta_{\bar{n}_i,\bar{n}_{P(i)}}=\det \,
\delta_{\bar{n}_i,\bar{n}_j}.
\end{equation}
The basic observation is now that, due to the symmetry of
$f(\bar{n}_1,...\bar{n}_N)$, only the conjugacy classes of $S_{N}$ are
relevant in computing the series: to see this we use the cycle decomposition
of the elements of $S_N$.

A conjugacy class of $S_N$ is conveniently described by the set of
non-negative
integers $\left\{\nu_i\right\}=(\nu_1,\nu_2,...,\nu_N)$
(we follow the description of \cite{libro}) satisfying the constraint
\begin{equation}
\nu_1+2\nu_2+3\nu_3+...+N\nu_N=N.
\end{equation}
Every element belonging to $\{\nu_i\}$ has the same parity and
can be decomposed, in a standard way, into $\nu_1$ one-cycles, $\nu_2$
two-cycles,..., $\nu_N$ $N$-cycles.
Due to the symmetry of $f(\bar n_1,...,\bar n_N)$ all
the elements of a conjugacy class give the same contribution in
eq.~(\ref{rapre}), as a simple relabeling of the $\bar{n}_i$'s and $\theta_i$'s
is sufficient: only the parity of the class and the
number of its elements, as function of $\{\nu_i\}$, are therefore relevant
to the computation of the partition function. It turns out
that $(-1)^{\sum_{i}\nu_{2i}}$ is the parity, while
the number of elements in the conjugacy class $\{\nu_i\}$ is
\begin{equation}
M_{\{\nu_i\}}=\frac{N!}{1^{\nu_1}\nu_1!\,2^{\nu_2}\nu_2!..N^{\nu_N}\nu_N!}.
\end{equation}
The next step is to use the decomposition in cycles
to perform explicitly the angular integrations: the effect is to express
the full series as a finite sum of series over a decreasing number of
integers. One easily realizes that a two-cycle results into the
identification of two $\bar{n}_i$'s in the sum, a three-cycle into the
identification of three $\bar{n}_i$'s and so on. We end up with
\begin{eqnarray}
&&{\cal Z}=
\displaystyle{\frac{(-1)^{(N-1)m}}{N!}\sqrt{\frac{2\pi}{g_c^2A_cN}}
\sum_{\{\nu_i\}}
\sum_{\bar n_1,..\bar n_\nu=-\infty}^{+\infty}(-1)^{\sum_{i}\nu_{2i}} M_{\{\nu_i\}}
\int_{-\infty}^{+\infty}\frac{d\beta}{2\pi}\,dy\int_{0}^{2\pi}
\frac{d\alpha}{\sqrt{\pi}}}
\nonumber\\
&&\displaystyle{\exp\left[-(\alpha-\frac{2\pi}{N}y)^2+i\beta
\left(y-\sum_{i_1=1}^{\nu_1}\bar n_{i_1}-
2\sum_{i_2=\nu_1+1}^{\nu_1+\nu_2}\bar n_{i_2}
-3\sum_{i_3=\nu_1+\nu_2+1}^{\nu_1+\nu_2+\nu_3}\bar n_{i_3}-...\right)+
2\pi i \frac{m}{N}y\right]}
\nonumber\\
&&\displaystyle{\exp\left[-\frac{g_c^2 A_c}{2}\left(\sum_{i_1=1}^{\nu_1}\bar n_{i_1}^2+
2\sum_{i_2=\nu_1+1}^{\nu_1+\nu_2}\bar n_{i_2}^2
+3\sum_{i_3=\nu_1+\nu_2+1}^{\nu_1+\nu_2+\nu_3}\bar n_{i_3}^2+...\right)-
\frac{g_c^2 A_c}{2}y^2\right]}
\label{partitione61}
\end{eqnarray}
where each conjugacy class has produced a sum over
$\nu=\nu_1+\nu_2+..+\nu_N$ integers; of course if some $\nu_j$ is zero, the
integers $n_{\nu_1+..\nu_{j-1}+1},..,n_{\nu_1+..\nu_{j-1}+\nu_{j}}$ do not
appear.
The Poisson resummation is, at this point, almost trivial, being the set
($n_1,..,n_\nu$) unrestricted: it requires in our case only gaussian
integrations. The remaining integrals over the auxiliary variables are also
easily done. The final result, expressing the original partition function as
a sum over ``dual'' integers $m_i$'s, is:
\begin{eqnarray}
&&{\cal Z}=
\displaystyle{\frac{(-1)^{(N-1)m}}{N}\sqrt{\frac{2\pi}{g_c^2A_cN}}
\left[1+{\sum_{\{\nu_i\}}}^{'}(-1)^{\sum_{i}\nu_{2i}}
Z_{\{\nu_i\}}(\frac{2\pi}{g_c^2A_c})^{\frac{\nu-1}{2}}\!\!\!\!\!\!\!\!\!\!
\sum_{m_1,..m_\nu=-\infty}^{+\infty}\delta(m-\sum_{i=1}^{\nu}m_i)\right.}\nonumber\\
&&\displaystyle{\left.
\exp\left[-\frac{2\pi^2}{g_c^2 A_c}\left(\sum_{i_1=1}^{\nu_1}
(m_{i_1}-\frac{m}{N})^2+\frac{1}{2}\sum_{i_2=\nu_1+1}^{\nu_1+\nu_2}
(m_{i_2}-2\frac{m}{N})^2+\frac{1}{3}\sum_{i_3=\nu_1+\nu_2+1}^{\nu_1+\nu_2+\nu_3}
(m_{i_2}-3\frac{m}{N})^2+ \right.\right.\right.}\nonumber\\
&&\displaystyle{\left.\left.\left. ..\frac{1}{N-1}
\sum_{i_{N-1}}^{\nu}(m_{i_{N-1}}-(N-1)\frac{m}{N})^2\right)\right]
\right]}
\label{partitione6}
\end{eqnarray}
where the prime means that we are summing over all the partitions
of $N$ except the identity, $\nu_1+2\nu_2+..+(N-1)\nu_{N-1}=N$ and
\begin{equation}
Z_{\{\nu_i\}}=\frac{N^{\frac{3}{2}}
\left[1^{\nu_1}2^{\nu_2}3^{\nu_3}...(N-1)^{\nu_{N-1}}\right]
^{-\frac{3}{2}}}{\nu_1!\nu_2!\nu_3!...\nu_{N-1}!}.
\end{equation}
These formulae have nice interpretation in term of instantons: the partition
function appears to be localized around its classical solutions \cite{witten}
and the limit $g_c^2A_c\to 0$ is related to the properties of the moduli
space of flat connections. Using eqs. (\ref{defi1}), (\ref{defi2}),
(\ref{defi3}) we can rewrite the sum in eq.(\ref{partitione6}) in the
following way (we consider the case of $\theta$ positive):
\begin{eqnarray}
&&\!\!\!\!\!\!\!\!
\displaystyle{{\sum_{\{\nu_i\}}}^{'}(-1)^{\sum_{i}\nu_{2i}}
Z_{\{\nu_i\}}(\frac{N^2}{g_{nc}^2\theta})^{\frac{\nu-1}{2}}\!\!\!\!\!
\sum_{m_1,..m_\nu=-\infty}^{+\infty}\!\!\!\!\!\!
\delta(\sum_{i=1}^{\nu}m_i+1) \exp\left[-\frac{\pi N^2}{g_{nc}^2
\theta}\left(\sum_{i_1=1}^{\nu_1}
(m_{i_1}+\frac{1}{N})^2+\right.\right.}\\ &&\!\!\!\!\!\!\!\!
\displaystyle{\left.\left.\frac{1}{2}\sum_{i_2=\nu_1+1}^{\nu_1+\nu_2}
(m_{i_2}+\frac{2}{N})^2+\frac{1}{3}\sum_{i_3=\nu_1+\nu_2+1}^{\nu_1+\nu_2+\nu_3}
(m_{i_3}+\frac{3}{N})^2+..\frac{1}{N-1}\sum_{i_{N-1}}^{\nu}
(m_{i_{N-1}}+\frac{N-1}{N})^2\right)\right]}.\nonumber
\label{partitione7}
\end{eqnarray}
At this point we have to perform our large $N$-limit: in so doing we are
recovering the theory on the non-commutative plane. We will neglect
exponentially suppressed contribution: in this way we expect to find the
partition function expressed through a semiclassical expansion, taking
into account all the finite action classical solutions. We do not consider
here the (possible) effect of non finite action contributions.

The problem is therefore reduced to find all the configurations that are not
exponentially
suppressed in the limit $N\to \infty$: we have to single out
both the partitions of $N$ and the instanton numbers $m_i$.
A very simple set surviving in the limit is the following: let us consider
the conjugacy classes composed by a single cycle of order $N$
and the other ones finite
\begin{equation}
\nu_{N-k}=1;\,\,\,\,\nu_1+2\nu_2+..+k\nu_{k}=k,
\label{party1}
\end{equation}
where $k$ is some positive integer. The choice
$m_{i_{1}}=m_{i_{2}}=..=m_{i_{k}}=0;\,m_{i_{N-k}}=-1$ for the associated
instanton numbers leads to a finite exponent
\begin{eqnarray}
&&\displaystyle{\frac{\pi N^2}{g_{nc}^2 \theta}
\left(\frac{\nu_1+2\nu_2+..+k\nu_{k}}{N^2}+
\frac{k^2}{N-k}\frac{1}{N^2}\right)}\nonumber\\
&&=\frac{\pi k}{g_{nc}^2 \theta}+O(\frac{1}{N}).
\end{eqnarray}
We claim that these are the dominant configurations: we do not have a rigorous proof but it is
possible to show, first of all,  that configurations with no cycle of order $N$ are suppressed.
Suppose to have a partition with cycles up to $k$ $$\nu_1+2\nu_2+..+k\nu_k=N.$$ Computing
the exponent and using the fact that $\sum_{i=1}^{\nu}m_i=-1$ we obtain:
\begin{equation}
\frac{\pi N^2}{g_{nc}^2 \theta}\left(\sum_{i_1}m_{i_{1}}^2+
\frac{1}{2}\sum_{i_2}m_{i_{2}}^2+..+\frac{1}{k}\sum_{i_k}m_{i_{k}}^2-
\frac{1}{N} \right);
\end{equation}
the contribution inside the parenthesis is $O(1)$. On the other hand when two cycles
are of order $N$ we get a suppressed configuration as well. Let us
consider for example the partition $N_1+N_2=N$ ($N_1<N_2$): because the ratios $N_1/N,N_2/N$ are
finite in the limit, we can minimize the action along $m_1+m_2=-1$ finding
$$\frac{\pi N^2}{g_{nc}^2 \theta}\left(\frac{1}{N_1}(m_1+\frac{N_1}{N})^2+
\frac{1}{N_2}(m_2+\frac{N_2}{N})^2\right)=\frac{\pi N_1}{g_{nc}^2 \theta}
(1+\frac{N_1}{N_2}).$$
It is easy to generalize the argument when finite-size cycles are also present and more than two
cycles of size $N$ appear in the decomposition. We assume therefore that the configurations
of eq.(\ref{party1}) are the dominant ones in the large $N$-limit: summing over $k$ we obtain
\begin{equation}
{\cal
Z}=\displaystyle{\frac{1}{N\sqrt{g_{nc}^2\theta}}\sum_{k=0}^{+\infty}
(-1)^k\exp(-\frac{\pi k}{g_{nc}^2\theta})\sum_{\{\nu_i\}_k}
\frac{(-1)^{\sum_{i}\nu_{2i}}}
{[2^{\nu_2}...k^{\nu_k}]^{\frac{3}{2}}}
\frac{1}{\nu_1!..\nu_k!}\left(\frac{A_{nc}}{\sqrt{2\pi
g_{nc}^2\theta^3}}\right)^\nu,} \label{partitione8}
\end{equation}
where we have substituted $N$ with the area of the non-commutative plane and
the explicit form
of $Z_{\{\nu_i\}}$ has been taken into account. This formula displays a
certain number of
remarkable features: probably the most interesting is that the partition
function is expressed
as a sum over ''fluxons''. In \cite{poly,gn2} it
has been shown that pure $U(1)$
gauge theory on the non-commutative plane admits finite energy instanton
solutions carrying
quantized magnetic flux. A peculiar feature is that fluxons exist with only
one sign of
the magnetic charge, $B$ being aligned with $\theta$. The classical action $S$
is also peculiar being linear in the magnetic number: it turns out that for a
fluxon solution of
charge $m$ we have
\begin{equation}
S=\frac{\pi m}{g^2\theta}.
\end{equation}
Consistently with the non-existence of classical solutions in the
commutative case, we see
that this action is singular as $\theta\to 0$.
Eq.(\ref{partitione8}) reproduces correctly
the sum over the classical action: one is therefore tempted to interpret the
coefficient associated to the charge $k$ sector as the effect of the
fluctuation around the fluxon
solution. According to \cite{poly,gn2}, a fluxon of charge $k$ carries a
$2k$-dimensional moduli space
reflecting the centers of the $k$ elementary vortices of which is made.
Moreover these solutions
have been proved to be unstable
\cite{gn2,andy} and a sensible semiclassical expansion would seem
therefore to be hopeless.
On the other hand it is well known that the partition function of ordinary
$YM_2$, on compact surface, is localized around its critical points
\cite{witten}, due to a generalization of the
Duistermaat-Heckman formula: the path integral is exact in the semiclassical approximation although the classical solutions are unstable.
In the non-commutative case, at least in the
limit of large non-commutative area, the very same phenomenon seems to occur: the coefficient
associated to $k$ has the following appealing explanation in terms of moduli.
If we assume
that an elementary vortex could carry ${\it any}$ integer charge, the sum over the partition of
$k$ in eq.(\ref{partitione8}) has a natural interpretation: a magnetic charge $k$ appears to
be composed by $\nu_1$ elementary vortex of charge
$1$, $\nu_2$ of charge $2$ and so on.
Vortices of equal charge, inside the fluxon, are identical, therefore the factor
$\frac{1}{\nu_1!\nu_2!..\nu_k!}$ appears. The integration over the positions is also correctly
reproduced, the area factor appearing with exponent $\nu_1+\nu_2..+\nu_k$, that represents the
total number of elementary constituents. The other factors have to be related to the computation
of the quantum fluctuations around the fluxons.
The solutions found in \cite{poly,gn2}, in this
picture, are the one related to $\{\nu_1=k,\,\,\nu_2=..=\nu_k=0\}$.
It could be that on compact
space a larger set of solutions (we remark that we are decompactifying a torus) is present
but we do not have an answer at this moment. Alternatively one could write the partition
function as $${\cal Z}=\sum_{k=0}^{+\infty}\exp(-\frac{\pi k}{g_{nc}^2\theta})
\frac{a_k(A_{nc}/\sqrt{g_{nc}\theta})}{k!},$$ and to understand $a_k$ as the total result of
the fluctuations around the Gross-Nekrasov solution. In any case eq.(\ref{partitione8}) may
provide an answer to the question posed in ref. \cite{gn2} about the possibility to
compute the non-commutative partition function (the regulator we use is essentially the area of
torus) in the semiclassical approximation. At this point we can actually go further: it is possible to resum exactly the series in
eq. (\ref{partitione8}). To compute the sum we observe that the constraint
$\nu_1+2\nu_2+...+k\nu_k=k$ can be implemented by introducing an angular
variable to obtain
\begin{eqnarray}
{\cal Z}&&=\sum_{k=0}^{+\infty}
(-1)^k\exp(-\frac{\pi k}{g_{nc}^2\theta})\sum_{\{\nu_i\}=0}^{+\infty}
\int_{0}^{2\pi}\frac{d\theta}{2\pi}
\exp\Biggl[\,i(\nu_1+2\nu_2+...+k\nu_k-k)\theta\Biggr]\nonumber\\
&&\left(\frac{A_{nc}}{\sqrt{2\pi g_{nc}^2\theta^3}}\right)^{\nu_1+\nu_2+..+\nu_k}
(-1)^{\nu_2+\nu_4+...}\,\,\,\frac{(1^{-\frac{3}{2}})^{\nu_1}}{\nu_1!}
\frac{(2^{-\frac{3}{2}})^{\nu_2}}{\nu_2!}...
\frac{(N^{-\frac{3}{2}})^{\nu_k}}{\nu_k!},
\label{titti}
\end{eqnarray}
where we disregarded the overall multiplicative constant.
The sum over $\nu_i$'s is now simple, giving
\begin{equation}
{\cal Z}=\sum_{k=0}^{+\infty}(-1)^k\exp(-\frac{\pi k}{g_{nc}^2\theta})
\int_{0}^{2\pi}\frac{d\theta}{2\pi}
{\rm e}^{-ik\theta}\exp\left[-\frac{A_{nc}}{\sqrt{2\pi g_{nc}^2\theta^3}}\sum_{l=1}^{k}
\frac{{\rm e}^{il\theta}(-1)^l}{l^{\frac{3}{2}}}\,\right],
\end{equation}
that can be expressed as a contour integral in the complex plane
\begin{equation}
{\cal Z}=\sum_{k=0}^{+\infty}(-1)^k\exp(-\frac{\pi k}{g_{nc}^2\theta})
\frac{1}{2\pi i}\int_{{\cal C}_0}\frac{dz}{z^{k+1}}
\exp\left[z\Phi(-z;\frac{3}{2};1)\frac{A_{nc}}{\sqrt{2\pi g_{nc}^2\theta^3}}\,\,\right],
\label{contorno}
\end{equation}
where ${\cal C}_0$ rounds the origin anticlockwise, sufficiently close so that
the function $\Phi(z;s;\mu)$
\begin{equation}
\Phi(z;s;\mu)=\sum_{k=0}^{+\infty}\frac{z^k}{(k+\mu)^s};
\end{equation}
is analytic ($|z|<1$ to avoid the cut from $z=1$ to $\infty$ of the $\Phi$ function). Now
${\cal C}_0$ can be chosen so that
$|\displaystyle{\frac{e^{-\frac{\pi}{g^2_{nc}\theta}}}{z}}|<1$ and the
series in $k$ is easily done
\begin{equation}
{\cal Z}=\frac{1}{2\pi i}\int_{{\cal C}_0}\frac{dz}{z+e^{-\frac{\pi}{g^2_{nc}\theta}}}
\exp\left[z\Phi(-z;\frac{3}{2};1)\frac{A_{nc}}{\sqrt{2\pi g_{nc}^2\theta^3}}\,\,\right],
\end{equation}
leading to
\begin{equation}
{\cal Z}=\exp\left[-\frac{e^{-\frac{\pi}{g^2_{nc}\theta}}}{\sqrt{2\pi g_{nc}^2\theta^3}}
\Phi(e^{-\frac{\pi}{g^2_{nc}\theta}};\frac{3}{2};1)A_{nc}\,\,\right].
\label{exte}
\end{equation}
The formula is interesting: the partition function, in our approximation,
appears to be extensive and the whole instantons series has been resummed into
the coefficient of $A_{nc}$ in eq. (\ref{exte}). This is typical of the
diluite instanton gas picture, in which instantons are taken to be
not interacting. In our case this result is exact, fluxons being non
interacting objects \cite{gn2}. We will use the expression of ${\cal Z}$ as
normalization for the correlation function of two open Wilson lines.

\section{The correlator of open Wilson Lines}

Now we would like to apply our machinery to the computation of some physical
observables: in a non-commutative gauge theory the basic gauge invariant
quantities are the correlators of open Wilson lines \cite{ghi}
(see also \cite{iikk}, where they were originally proposed, and \cite{dr}
concerning their supergravity description). Pure
non-commutative gauge theory has no local gauge invariant operators: it
is instead possible to construct a gauge invariant observable out of the
open Wilson lines provided they have a transverse momentum (we will consider
for simplicity straight open Wilson lines)
\begin{equation}
W(\vec p)=\int d^2x\,\Omega(x_i\to x_i+\theta_{ij}p_j)
\star \exp(i\vec p \,\vec x),
\label{wl1}
\end{equation}
where $\Omega(x\to y)$ is an open Wilson line, path ordered with respect
to the non-commutative star product, and stretching between the point $x$ and
$y$ in the noncommutative plane. Because of the noncommutativity, the term
$\exp(i\vec p\,\vec x)$ is a translation operator in the direction
transverse to $\vec p$ which for gauge invariance must relate the two
endpoints of the open Wilson line \cite{ghi}. Quantum correlators were
also considered in \cite{ghi,dk}, exhibiting an interesting string-like
behavior.
We are interested here in computing the simplest non-trivial correlator (the
two-point function) for straight, parallel Wilson lines. We will obtain a
finite result, in the limit of long lines,
expressed through an expansion around the classical solutions
of the theory, in agreement with the result for the partition function.

We start by considering the following observable, stretching along $x_2$,
on the noncommutative torus:
\begin{equation}
W(k_1,n_2)=\frac{1}{4\pi^2 R^2_{nc}}
\int_0^{2\pi R_{nc}} d^2x\,\Omega(x_1;x_2\to x_2+2\pi R_{nc}n_2
-2\pi R_{nc}\frac{c}{N}k_1)\star \exp(i\frac{k_1 x}{R_{nc}}).
\label{wl2}
\end{equation}
This is the direct generalization of eq. (\ref{wl1}):
$n_2$ describes the number of winding along the $x_2$ coordinate and
$k_1$ is the integer
associated to the transverse momentum $k_1/R_{nc}$. In order to
count into $n_2$ all the windings we take $|ck_1|<N$. The normalization
of the line is chosen so that $W(0,0)=1$. Let us consider the
decompactification limit in eq. (\ref{wl2}) (necessarily implying $n_2=0$),
requiring to have a finite transverse momentum,
$\displaystyle{\frac{k_1}{R_{nc}}\to p}$ as $R_{nc}\to \infty$.
Taking into account eq. (\ref{dec}) we see that $k_1$ has to be scaled as
\begin{equation}
k_1=\sqrt{\frac{N|\theta|}{2\pi}}\,p.
\end{equation}
In the decompactification limit the total length $L$
of the Wilson line turns out to be finite
\begin{equation}
L=\frac{2\pi R_{nc}k_1}{N}=p|\theta|,
\label{lunghezza}
\end{equation}
with the only restriction $p\sqrt{\frac{\theta}{2\pi}}<\sqrt{N}$
(that is the original no winding request).

The next step is to write down explicitly the Morita transformed operator:
the mapping has been derived in \cite{gura,amns,sara}, showing that open
non-commutative Wilson lines map into the Polyakov loops of ordinary
Yang-Mills theory,
\begin{eqnarray}
W(k_1,0)&&\displaystyle{=W^{(k_1)}}\nonumber\\
W^{(k_1)}&&\displaystyle{=\frac{1}{4\pi^2R^2_{c}}\int_0^{2\pi R_{c}}
d^2x\,\frac{1}{N}{\rm Tr}\,\left[\,\Omega^{(k_1)}(x_1)\,\right].}
\label{wl3}
\end{eqnarray}
$\Omega^{(k_1)}(x_1)$ is the holonomy derived from a closed path, winding
$k_1$ times along the $x_2$ direction, the trace in eq. (\ref{wl3}) has to
be taken  in the fundamental representation of $U(N)$.
We see therefore that under Morita equivalence the computation of
open Wilson lines correlators has been mapped in an analogous problem for
conventional Polyakov loops. As for the classical action, we have
to consider in
the definition of $\Omega^{(k_1)}(x_1)$ the contribution of the background
connection eq.(\ref{beck}):
it turns out that the holonomy of the fixed abelian
background has to be subtracted, leaving us with the computation of
$\Omega^{(k_1)}(x_1)$ for $U(1)\otimes SU(N)/Z\!\!\!Z_N$
(in the flux sector $m_c$), where the $U(1)$ contribution is taken in the
trivial sector. Let us consider the simplest correlation function,
involving just two open Wilson lines \cite{ghi},
\begin{equation}
W_2(k_1)=<W(k_1,0)W(-k_1,0)>:
\end{equation}
under Morita equivalence we have
\begin{equation}
W_2(k_1)=\frac{1}{4\pi^2 R_c^2}\int_0^{2\pi R_c} dx_1 dy_1
<\!\!\frac{1}{N}{\rm Tr}\,\left[\,\Omega^{(k_1)}(x_1)\,\right]
\frac{1}{N}{\rm Tr}\,\left[\,\Omega^{(-k_1)}(y_1)\,\right]\!\!>,
\label{wl41}
\end{equation}
where the integrations over $x_2,y_2$ lead simply to a volume factor because
the correlator does not depend on them. We can factorize another volume
factor by noticing that due to translational invariance the correlation
function only depends on the relative coordinate and distinguishing the
physically inequivalent configurations we arrive to
\begin{equation}
W_2(k_1)=\frac{1}{2\pi R_c}\left[\int_0^{\pi R_c} dx\,
<\frac{1}{N}{\rm Tr}\,\left[\,\Omega^{(k_1)}(x)\,\right]
\frac{1}{N}{\rm Tr}\,\left[\,\Omega^{(-k_1)}(0)\,\right]>+\,k_1\to-k_1\right].
\label{wl4}
\end{equation}
The correlation function has to be computed in the $m$-th 't Hooft sector
(we will consider at
the end $m=\pm 1$ according to the sign of $\theta$ as we have done for the
partition function)
and
we can again take
advantage of the Migdal-Rusakov's formulae for $U(N)$:
\begin{eqnarray}
&&\frac{1}{N^2}<\!{\rm Tr}\,\left[\,\Omega^{(k_1)}(x)\,\right]\,{\rm Tr}\,
\left[\,\Omega^{(-k_1)}(0)\,\right]\!>=\frac{1}{{\cal Z}}
\sum_{R,S} \exp\left[-{{g_c^2 (A_c-A_2)}\over 2}C_2(R)\right]\times\nonumber\\
&&\!\!\!\!\!\!\!\!\!\!\!\!\!\!\!\!
\exp\left[-{{g_c^2 A_2}\over 2}C_2(S)\right]
\int dU_1\,{\cal X}_R(U_1){\cal X}_F(U_1^{k_1}){\cal X}_S^\dagger(U_1)
\int dU_2\,{\cal X}_S(U_2){\cal X}_F(U_2^{-k_1}){\cal X}_R^\dagger(U_2),
\end{eqnarray}
$A_2$ being the total area contained between the two loops ($A_2=2\pi R_cx$)
and  ${\cal X}_F$ is the character of the
fundamental representation of $U(N)$.
It is not difficult single out the relevant
$U(1)\otimes SU(N)/Z\!\!\!Z_N$ contribution, along the same lines of the
previous chapter. In term of the
$\bar{n}_i$'s we have the analogue of eq. (\ref{partitione4})
\begin{eqnarray}
&&\displaystyle{\frac{1}{N^2}<\!{\rm Tr}\,\left[\,\Omega^{(k_1)}(x)\,\right]\,
{\rm Tr}\,\left[\,\Omega^{(-k_1)}(0)\,\right]\!>=
\frac{1}{{\cal Z}N!}\sqrt{\frac{2\pi}{g_c^2A_cN}}
\exp\left[\frac{g^2_cA_2^2k_1^2}{2A_cN}-\frac{g^2_c A_2 k_1^2}{2}\right]}
\nonumber\\
&&\displaystyle{\sum_{\bar{n}_i\neq \bar{n}_j}
\int_{0}^{2\pi}\frac{d\alpha}{\sqrt{\pi}}
\exp\left[-(\alpha-\frac{2\pi}{N}\sum_{i=1}^{N}\bar{n}_i)^2-2\pi i \frac{m}{N}
\left(\frac{(N-1)N}{2}-\sum_{i=1}^{N}\bar{n}_i\right)\right]}\nonumber\\
&&\exp\left[-\frac{g_c^2 A_c}{2}\left(\sum_{i=1}^{N}\bar{n}_{i}^2-\frac{1}{N}
(\sum_{i=1}^{N}\bar{n}_i)^2\right)\right]\frac{1}{N}\sum_{j=1}^{N}
\exp\left[-g_c^2 A_2 k_1(\bar{n}_j-\frac{1}{N}
\sum_{i=1}^{N}\bar{n}_{i}) \right].
\label{wl5}
\end{eqnarray}
We will take for simplicity, from now on, $N$ odd. We introduce the cycles
decomposition, as in eq. (\ref{partitione61}), and, using the auxiliary
integration variables, we can write the sum in eq. (\ref{wl5}) as
\begin{eqnarray}
&&\displaystyle{\frac{1}{N}
\sum_{\{\nu_i\}}
\sum_{\bar n_1,..\bar n_\nu=-\infty}^{+\infty}(-1)^{\sum_{i}\nu_{2i}}
M_{\{\nu_i\}}
\int_{-\infty}^{+\infty}\frac{d\beta}{2\pi}\,dy\int_{0}^{2\pi}
\frac{d\alpha}{\sqrt{\pi}}}\exp\left[-\left(\alpha-\frac{2\pi}
{N}y\right)^2\right]\nonumber\\
&&\displaystyle{\exp\left[-\frac{g_c^2 A_c}{2}
\left(\sum_{i_1=1}^{\nu_1}\bar n_{i_1}^2+
2\sum_{i_2=\nu_1+1}^{\nu_1+\nu_2}\bar n_{i_2}^2
+3\sum_{i_3=\nu_1+\nu_2+1}^{\nu_1+\nu_2+\nu_3}\bar n_{i_3}^2+...\right)-
\frac{g_c^2 A_c}{2}y^2\right]}\nonumber\\
&&\displaystyle{\exp\left[i\beta
\left(y-\sum_{i_1=1}^{\nu_1}\bar n_{i_1}-
2\sum_{i_2=\nu_1+1}^{\nu_1+\nu_2}\bar n_{i_2}
-3\sum_{i_3=\nu_1+\nu_2+1}^{\nu_1+\nu_2+\nu_3}\bar n_{i_3}-...\right)
+\frac{2\pi iy}{N}
\left(m-i\frac{g^2_cA_2k_1}{2\pi}\right)
\right]}
\nonumber\\
&&\displaystyle{\,\,\,\,\,\,\,\,\,\,\,
\Bigl[\nu_1\exp\left(-g^2_cA_2k_1n_1\right)
+2\nu_2\exp\left(-g^2_cA_2k_1n_{\nu_1+1}\right)
+...\,\,\Bigr]},
\label{wl6}
\end{eqnarray}
where we have explicitly taken into account the symmetry between the integers
associated to cycles of equal length. The Poisson resummation is now simple
because we can use our previous result realizing that we have the shifts
\begin{eqnarray}
&&m\to m-i\frac{g^2_cA_2k_1}{2\pi},\nonumber\\
&&m_1\to m_1-i\frac{g^2_cA_2k_1}{2\pi},\,\,\,
m_{\nu_1+1}\to m_{\nu_1+1}-i\frac{g^2_cA_2k_1}{2\pi},\,\,.....
\end{eqnarray}
The final result can be written as
\begin{eqnarray}
&&\displaystyle{\frac{1}{N^2}<\!{\rm Tr}\,\left[\,\Omega^{(k_1)}(x)\,\right]\,
{\rm Tr}\,\left[\,\Omega^{(-k_1)}(0)\,\right]\!>=
\frac{1}{{\cal Z}N}\sqrt{\frac{2\pi}{g_c^2A_cN}}
\exp\left[\frac{g^2_cA_2^2k_1^2}{2A_cN}-\frac{g^2_c A_2 k_1^2}{2}\right]}
\Biggl\{1+
\nonumber\\
&&\displaystyle{
{\sum_{\{\nu_i\}}}^{'}(-1)^{\sum_{i}\nu_{2i}}
Z_{\{\nu_i\}}(\frac{2\pi}{g_c^2A_c})^{\frac{\nu-1}{2}}\!\!\!\!\!\!\!\!\!\!
\sum_{m_1,..m_\nu=-\infty}^{+\infty}\delta(m-\sum_{i=1}^{\nu}m_i)
\exp\left[-\frac{2\pi^2}{g_c^2 A_c}\left(\sum_{i_1=1}^{\nu_1}
(m_{i_1}-\frac{m}{N})^2+\right.\right.}
\nonumber\\
&&\displaystyle{\left.\left.\left.
+\frac{1}{2}\sum_{i_2=\nu_1+1}^{\nu_1+\nu_2}
(m_{i_2}-2\frac{m}{N})^2+....\right)\right]\frac{1}{N}\sum_{l=1}^{N-1}l\nu_l
\exp\left[\frac{g^2_cA_2^2k_1^2}{2A_c}
\left(\frac{1}{l}+\frac{l}{N^2}-\frac{2}{N}\right)+\right.\right.}\nonumber\\
&&\displaystyle{\left.\left.
+2\pi i\frac{A_2k_1}{A_c}\left(\frac{1}{l}-\frac{1}{N}\right)
\left(m_{\nu_1+..+\nu_{l-1}+1}-\frac{ml}{N}\right)
\right]\right\}.}
\end{eqnarray}
Next we have to evaluate the above expression on the configurations found
in the previous section and to take the large $N$-limit: we arrive to the
following expression (we write the sum in terms of the non-commutative data)
\begin{eqnarray}
1\,\,+
&&\displaystyle{\frac{1}{{\cal Z}N\sqrt{g^2_{nc}\theta}}\sum_{k=1}^{+\infty}
(-1)^k\exp(-\frac{\pi k}{g_{nc}^2\theta}){\sum_{\{\nu_k\}}}^{'}
(-1)^{\sum_i\nu_{2i}}
Z_{\{\nu_k\}}\left(\frac{A_{nc}}{\sqrt{2\pi g_{nc}^2\theta^3}}\right)^\nu
}\nonumber\\
&&\displaystyle{\frac{1}{N}\left(\sum_{l=1}^k l\nu_l\exp\left[
i\frac{xp}{2\pi}+\frac{1}{l}\frac{g_{nc}^2p^2x^2\theta}{16\pi^3}\right]
-k\right)}.
\end{eqnarray}
In order to evaluate the sum let us discuss
$$
\displaystyle{F(xp,y)=\sum_{k=1}^{+\infty}
(-1)^k\exp(-\frac{\pi k}{g_{nc}^2\theta}){\sum_{\{\nu_k\}}}^{'}
(-1)^{\sum_i\nu_{2i}}
Z_{\{\nu_k\}}y^\nu\frac{1}{N}\sum_{l=1}^k l\nu_lf_l(xp)},
$$
where $\displaystyle{y=\frac{A_{nc}}{\sqrt{2\pi g_{nc}^2\theta^3}}}$ and
$\displaystyle{f_l(xp)=\exp\left[i\frac{xp}{2\pi}+\frac{1}{l}
\frac{g_{nc}^2p^2x^2\theta}{16\pi^3}\right]}$: the correlator is
now simply given by
$$
\displaystyle{\frac{1}{N^2}<\!{\rm Tr}\,\left[\,\Omega^{(k_1)}(x)\,\right]\,
{\rm Tr}\,\left[\,\Omega^{(-k_1)}(0)\,\right]\!>=
1+\frac{1}{{\cal Z}N\sqrt{g^2_{nc}\theta}}\left[F(xp,y)-F(0,y)\right]}.
$$
Using the contour representation in eq. (\ref{contorno}) we have
\begin{eqnarray}
&&F(xp,y)=\displaystyle{\sum_{k=1}^{+\infty}(-1)^k
\exp(-\frac{\pi k}{g^2_{nc}\theta})\frac{y}{N}
\sum_{l=1}^kf_l(xp)\frac{(-1)^l}{\sqrt{l}}\frac{1}{2\pi i}
\int_{{\cal C}_0}\frac{dz}{z^{k+1}}z^l
\exp\left[z\Phi(-z;\frac{3}{2};1)\,y\,\,\right]}\nonumber\\
&&=\displaystyle{\sum_{k=0}^{+\infty}(-1)^k
\exp(-\frac{\pi k}{g^2_{nc}\theta})\frac{y}{N}
\sum_{l=1}^{+\infty}f_l(xp)\frac{(-1)^l}{\sqrt{l}}\frac{1}{2\pi i}
\int_{{\cal C}_0}\frac{dz}{z^{k+1}}z^l
\exp\left[z\Phi(-z;\frac{3}{2};1)\,y\,\,\right]}.
\label{contorno2}
\end{eqnarray}
Performing the geometric series in $k$ and evaluating the
contour integrals on the pole we obtain
\begin{equation}
\displaystyle{F(xp,y)=
-{\cal Z}N\sqrt{2\pi}\sum_{l=1}^{+\infty}
f_l(xp)\frac{\exp(-\frac{\pi l}{g^2_{nc}\theta})}{\sqrt{l}}}.
\end{equation}
The partition function has been explicitly factorized out leaving us with
the result
\begin{equation}
\displaystyle{\frac{1}{N^2}<\!{\rm Tr}\,\left[\,\Omega^{(k_1)}(x)\,\right]\,
{\rm Tr}\,\left[\,\Omega^{(-k_1)}(0)\,\right]\!>= 1+
\sqrt{\frac{2\pi}{g^2_{nc}\theta}}\sum_{l=1}^{+\infty}
\frac{\exp(-\frac{\pi l}{g^2_{nc}\theta})}{\sqrt{l}}\left(1-f_l(xp)\right)}.
\label{wl10}
\end{equation}
Let us notice that because $f_l(0)=1$ we still have that for vanishing length
the correlator is normalized to one. In terms of the length of the lines
the result reads
\begin{equation}
\displaystyle{1+\sqrt{\frac{2\pi}{g^2_{nc}\theta}}\sum_{l=1}^{+\infty}
\frac{\exp(-\frac{\pi l}{g^2_{nc}\theta})}{\sqrt{l}}\left(1-
\exp\left[i\frac{Lx}{\theta}+\frac{1}{l}\frac{g^2_{nc}L^2x^2}{16\pi^3\theta}
\right]\right)}.
\end{equation}
This is the unintegrated expression: it seems still to have an interpretation
in terms of fluxons. The phase is reminiscent of the result of \cite{gn2},
where the classical Wilson loop, evaluated on fluxons, has exactly the same
form ($Lx$ is the area determined by lines), not depending on the
coupling constant $g^2_{nc}$ nor on the fluxon charge. The other exponential
term has a peculiar dependence $\displaystyle{\frac{1}{l}}$ that calls for an
explanation: it would be naturally identified as the contribution of the
fluctuations around the classical fluxon and it would be important to see if
such a strange dependence could appear directly from some computations on
the non-commutative plane. The physical observable, nevertheless, is
the integral of eq. (\ref{wl10})
that, because $R_c\to 0$ as $\frac{1}{\sqrt{N}}$, seems to go to 1
(see eq. (\ref{wl4})).
Actually we can obtain a non-trivial result, in our approximation,
considering Wilson lines that are very long, or, if you want, with very
high momentum $p$. Coming back to the definition of $L$ in
eq.(\ref{lunghezza}), we see that we can consistently consider Wilson lines of
length
$$
L=\frac{\lambda}{\pi}R_{nc},
$$
with
arbitrary $0<\lambda<1/2$: taking into account the scaling of $L$ we
obtain the correlator
\begin{equation}
W_2(\lambda)=\displaystyle{
1+\frac{1}{2}\int_{-1}^1 dz \sqrt{\frac{2\pi}{g^2_{nc}\theta}}
\sum_{l=1}^{+\infty}\frac{\exp(-\frac{\pi l}{g^2_{nc}\theta})}{\sqrt{l}}\left(1-
\exp\left[i\lambda z+\frac{1}{l}\frac{g^2_{nc}\theta\lambda^2
z^2}{4\pi}
\right]\right)}.
\end{equation}
The fact that only for long Wilson lines our computation leads
to a non-trivial result may be related to point-like character of
the coupling between Wilson lines and fluxons (see the discussion in
\cite{gn2}) and therefore only when the line is enough long the
interaction becomes important. In any case we do not have a satisfactory
argument to support this thesis and it might happen that non-parallel
Wilson lines be non-trivial or one had to compute higher order correlators.

\section{Conclusions}
We have explored the possibility to study dynamical properties of
non-commutative gauge theories using the powerful tool of Morita
equivalence. We have restricted our attention to the two-dimensional case,
being the simplest situation in which concrete computations can be
performed. When formulated on a two-torus with a rational
non-commutativity parameter $\theta$, the $U(1)$ gauge theory maps, under
a Morita transformation, into a usual $U(N)$ theory in a given t' Hooft
sector. We have shown that the non-commutative theory on the plane with
arbitrary $\theta$ parameter can be obtained by a suitable
decompactification limit, involving a series of rational approximants.
Morita equivalence translates this procedure into a non-standard large
$N$-limit in the (dual) commutative $U(N)$ theory: as $N$ goes to infinity
not only the coupling constant scales with $1/N$ but also the commutative
torus shrinks to zero-size in the same way. The t'Hooft flux has
to be taken fixed. Starting from the exact Migdal-Rusakov solution for
Yang-Mills theory on the torus, we have been able to perform such a limit
on the partition function, by going to a dual representation obtained
from the Poisson resummation of the original series over the Young tableaux
integers. We have seen that there are finite action configurations
surviving in the limit (being not exponentially suppressed as $N$ goes to
infinity) and the partition function appears therefore localized around
them. This is probably our most interesting result: these finite action
configurations are in correspondence with the classical solutions found
Polychronakos and by Gross and Nekrasov on the non-commutative plane
\cite{poly,gn2} and
the partition function we obtain has a precise interpretation as a
semiclassical expansion around them. The whole series can
actually be resummed, leading to a result that is extensive in the
area: our expression closely resembles a dilute instanton gas approximation, the
Gross-Nekrasov fluxons being indeed not interacting. We have then shown
how to compute Wilson lines correlators: we have carried out the simple
case of two parallel Wilson lines. A non-trivial result, having an
instanton interpretation, has been obtained in the long lines-high momentum
limit. We think that our computations are a first step towards the
possibility to solve completely the theory, on the non-commutative
plane, having reduced the problem, in principle, to a sort of
large $N$-small area limit of the usual Yang-Mills theory on a torus.
Many aspects remain to be explored: first of all we have assumed that
the resulting theory on the plane does not depend on the particular
series of rational approximants, a fact that has to be checked (see
\cite{luis,gura} for a discussion at finite volume). While the generalization to
a nonabelian non-commutative theory seems not difficult (we have studied
the U(1) case with vanishing non-commutative Chern class),
it remains open the possibility to consider general fluxes on the
non-commutative side and then to sum over them, possibly with some instanton
angle. Moreover we have not considered, in the spirit of the semiclassical
approximation, the possibility that infinite action configurations might
be relevant on the plane, once resummed. On the other hand, Witten has
shown that ordinary Yang-Mills theory on compact surface is localized
around its classical solutions, therefore our result could be something
more than a semiclassical approximation. It would be very important,
in this sense, to have some computations (hopefully nonperturbative) directly
on the non-commutative plane for Wilson line correlators and to check it against
the calculation done along our procedure. The relation between our finite action
configurations and the exact solitons on non-commutative tori, found in
\cite{Kra1} and \cite{rief,cinesi} may also shed some light
on the mathematical structure beyond the limit.

Finally closed Wilson loop could also be studied: in \cite{bnt} a perturbative
computation has shown that interesting features
concerning smoothness in $\theta$ and large $\theta$-limit can be drawn from Wilson
loops analysis. We will present, in a forthcoming paper \cite{gsv}, the result of our
investigations on Wilson loops and how the procedure we have presented here is
consistent, in a particular limit, with the resummation of the perturbative
series on the non-commutative plane.

\section{Acknowledgements}
It is a pleasure to acknowledge useful discussions with Jose Fernandez Barbon,
Francesco Bonechi, Antonio Bassetto, Marco Billo', Alessandro D'Adda, Fedele
Lizzi, Marco Tarlini, Paolo Provero and Federica Vian.
A special thanks goes to Thomas Krajewski for  illuminating discussions and to
Richard Szabo for a clarifying correspondence. We also
thank the Cern Theory Division, where part of the work has been done, for
the warm hospitality and the financial support.


\begin{thebibliography}{999}
\bibitem{cds}A.~Connes, M.~R.~Douglas and A.~Schwarz,
JHEP {\bf 9802} (1998) 003 [hep-th/9711162];
 M.~R.~Douglas and C.~M.~Hull,
JHEP {\bf 9802} (1998) 008
[hep-th/9711165].
\bibitem{sw} N.~Seiberg and E.~Witten,
JHEP {\bf 9909} (1999) 032
[hep-th/9908142].
\bibitem{mrs} S.~Minwalla, M.~Van Raamsdonk and N.~Seiberg,
JHEP {\bf 0002} (2000) 020
[hep-th/9912072].
\bibitem{gn}
D.~J.~Gross and N.~A.~Nekrasov,
JHEP {\bf 0007} (2000) 034
[hep-th/0005204].
\bibitem{suski}A.~Matusis, L.~Susskind and N.~Toumbas,
JHEP {\bf 0012} (2000) 002
[hep-th/0002075].
M.~Hayakawa,
Phys.\ Lett.\ B {\bf 478} (2000) 394
[hep-th/9912094].
\bibitem{gp}L.~Griguolo and M.~Pietroni,
JHEP {\bf 0105} (2001) 032
[hep-th/0104217].
\bibitem{espe}F.~R.~Ruiz,
Phys.\ Lett.\ B {\bf 502} (2001) 274
[hep-th/0012171];

K.~Landsteiner, E.~Lopez and M.~H.~Tytgat,
JHEP {\bf 0106} (2001) 055
[hep-th/0104133];

M.~V.~Raamsdonk,
``The meaning of infrared singularities in noncommutative gauge theories''
[hep-th/0110093];

A.~Armoni and E.~Lopez,
``UV/IR mixing via closed strings and tachyonic instabilities''
[hep-th/0110113].
\bibitem{ns}N. A.~Nekrasov and A.~Schwarz,
Commun.\ Math.\ Phys.\  {\bf 198} (1998) 689
[hep-th/9802068].
\bibitem{valya}T.~J.~Hollowood, V.~V.~Khoze and G.~Travaglini,
JHEP {\bf 0105} (2001) 051 [hep-th/0102045].
\bibitem{schwarz}A.~Schwarz,
Nucl.\ Phys.\ B {\bf 534} (1998) 720
[hep-th/9805034].
\bibitem{dual}B.~Pioline and A.~Schwarz,
JHEP {\bf 9908} (1999) 021
[hep-th/9908019]
\bibitem{baal}G.~'t Hooft,
Nucl.\ Phys.\ B {\bf 153} (1979) 141.
\bibitem{boris}S.~Elitzur, B.~Pioline and E.~Rabinovici,
JHEP {\bf 0010} (2000) 011
[hep-th/0009009].
\bibitem{gura} Z.~Guralnik and J.~Troost,
JHEP {\bf 0105} (2001) 022
[hep-th/0103168].
\bibitem{luis}
L.~Alvarez-Gaume and J.~L.~Barbon,
``Morita Duality and Large-N Limits'' [hep-th/0109176].
\bibitem{Lizzi} For a rigorous mathematical approach to the problem in the
case of scalar theory, see G.~Landi, F.~Lizzi and R.~J.~Szabo,
Commun.\ Math.\ Phys.\  {\bf 217}, 181 (2001)
[hep-th/9912130].
\bibitem{migdal}
A.~A.~Migdal,
Sov.\ Phys.\ JETP {\bf 42} (1975) 413;
B.~E.~Rusakov,
Mod.\ Phys.\ Lett.\ A {\bf 5} (1990) 693.
\bibitem{poly} A.~Polychronakos, Phys. Lett. B {\bf 495} (2000) 407
[hep-th/0007043]
\bibitem{gn2}D.~J.~Gross and N.~A.~Nekrasov,
JHEP {\bf 0103} (2001) 044
[hep-th/0010090].
\bibitem{bnt}A.~Bassetto, G.~Nardelli and A.~Torrielli,
``Perturbative Wilson loop in two-dimensional non-commutative Yang-Mills
theory'' [hep-th/0107147].
\bibitem{gura2}
Z.~Guralnik,
``Strong coupling phenomena on the noncommutative plane'' [hep-th/0109079].
\bibitem{transi}R.~Forman,
Commun.\ Math.\ Phys.\  {\bf 151} (1993) 39.
\bibitem{witten} E.~Witten,
J.\ Geom.\ Phys.\  {\bf 9}, 303 (1992)
[hep-th/9204083].
\bibitem{gsv} L. Griguolo, D. Seminara and P. Valtancoli in preparation.
\bibitem{bigatti} D.~Bigatti,
``Gauge theory on the fuzzy torus'' [hep-th/0109018].
\bibitem{w4} D.~J.~Gross and W.~I.~Taylor,
Nucl.\ Phys.\ B {\bf 400}, 181 (1993)
[hep-th/9301068];
Nucl.\ Phys.\ B {\bf 403}, 395 (1993)
[hep-th/9303046].
\bibitem{DK} M.~R.~Douglas and V.~A.~Kazakov,
Phys.\ Lett.\ B {\bf 319}, 219 (1993)
[hep-th/9305047].
\bibitem{ruud} R.~E.~Rudd,
``The String partition function for QCD on the torus,''
[hep-th/9407176].
\bibitem{bgv}A.~Bassetto, L.~Griguolo and F.~Vian,
Annals Phys.\  {\bf 285} (2000) 185
[hep-th/0002093].
\bibitem{toro} L.~Griguolo,
Nucl.\ Phys.\ B {\bf 547}, 375 (1999)
[hep-th/9811050].
\bibitem{libro} W. Miller Jr.,{\it Symmetry groups and their applications},
Accademic Press, New York and London (1972).
\bibitem{andy}M.~Aganagic, R.~Gopakumar, S.~Minwalla and A.~Strominger,
JHEP {\bf 0104}, 001 (2001)
[hep-th/0009142].
\bibitem{ghi} D.~J.~Gross, A.~Hashimoto and N.~Itzhaki,
``Observables of non-commutative gauge theories'' [hep-th/0008075].
\bibitem{iikk} N.~Ishibashi, S.~Iso, H.~Kawai and Y.~Kitazawa,
Nucl.\ Phys.\ B {\bf 573} (2000) 573
[hep-th/9910004].
\bibitem{dr}S.~R.~Das and S.~Rey,
Nucl.\ Phys.\ B {\bf 590} (2000) 453
[hep-th/0008042].
\bibitem{dk}A.~Dhar and Y.~Kitazawa,
JHEP {\bf 0102} (2001) 004
[hep-th/0012170].
\bibitem{amns}J.~Ambjorn, Y.~M.~Makeenko, J.~Nishimura and R.~J.~Szabo,
JHEP {\bf 0005} (2000) 023
[hep-th/0004147].
\bibitem{sara}K.~Saraikin,
J.\ Exp.\ Theor.\ Phys.\  {\bf 91} (2000) 653
[Zh.\ Eksp.\ Teor.\ Fiz.\  {\bf 91} (2000) 755]
[hep-th/0005138].
\bibitem{Kra1}T.~Krajewski and M.~Schnabl,
JHEP {\bf 0108} (2001) 002
[hep-th/0104090]
\bibitem{rief}M.~A.~Rieffel,
J.\ Diff.\ Geom.\  {\bf 31} (1990) 535.
\bibitem{cinesi}B.~Y.~Hou, D.~T.~Peng, K.~J.~Shi and R.~H.~Yue,
``Solitons on noncommutative torus as elliptic algebras and elliptic models''
[hep-th/0110122].
\end{thebibliography}
\end{document}